\documentclass[aps,prl,twocolumn]{revtex4-1}
\usepackage{graphicx}
\usepackage{times}
\usepackage{amsmath,amssymb}
\usepackage{bm}

\usepackage{color}
\newcommand{\beq}{\begin{equation}}
\newcommand{\eeq}{\end{equation}}
\newcommand{\beqa}{\begin{eqnarray}}
\newcommand{\eeqa}{\end{eqnarray}}

\begin{document}

\title{{ Gaussian} 
quantum fluctuations in the superfluid-Mott phase transition}
\author{M. Faccioli} 
\affiliation{Dipartimento di Fisica e Astronomia ``Galileo Galilei'', 
Universit\`a di Padova, Via Marzolo 8, 35131 Padova, Italy}
\author{L. Salasnich}
\affiliation{Dipartimento di Fisica e Astronomia ``Galileo Galilei'', 
Universit\`a di Padova, Via Marzolo 8, 35131 Padova, Italy}
\affiliation{Istituto Nazionale di Ottica (INO) del Consiglio Nazionale 
delle Ricerche (CNR), \\ Via Nello Carrara 1, 50019 Sesto Fiorentino, Italy}

\begin{abstract}
Recent advances in cooling techniques make now possible 
the experimental study of quantum phase transitions, which are 
transitions near absolute zero temperature 
accessed by varying a control parameter. 
A paradigmatic example is the superfluid-Mott transition 
of interacting bosons on a periodic lattice. 
From the relativistic Ginzburg-Landau action of this 
superfluid-Mott transition we derive the 
elementary excitations of the bosonic system, 
which contain in the superfluid phase 
a gapped Higgs mode and a gappless Goldstone mode. We show that 
this energy spectrum is in good agreement with the available 
experimental data and we use it to extract, with the help of 
dimensional regularization, meaningful analytical formulas for the 
beyond-mean-field equation of state in two and three 
spatial dimensions. We find that, while the mean-field equation of 
state always gives a second-order quantum phase transition, 
the inclusion of { Gaussian} quantum 
fluctuations can induce a first-order quantum phase transition. This 
prediction is a strong benchmark for next future experiments 
on quantum phase transitions.
\end{abstract}

\maketitle

\newpage

The Bose-Hubbard model of interacting bosons on a periodic lattice 
was introduced in 1963 by Gersch and Knollman 
\cite{Gersch} to describe the coherent properties of 
granular superconductors. The model gained 
much success by the late 1980s \cite{Ma,Giamarchi,Fisher}.  
More recently it has been used to investigate 
superconductivity and ultracold atoms in optical 
lattices \cite{Bruder,Sachdev}, 
but also quantum information \cite{Romero} and 
quantum chaos \cite{Kolowsky}. The bosonic gas described 
the Bose-Hubbard model displays 
a quantum phase transition between a superfluid 
phase and the Mott insulating phase 
\cite{Ma,Giamarchi,Fisher,Sachdev,Greiner,Kacz}. This transition 
corresponds to a global $U(1)$ spontaneous symmetry breaking 
\cite{Goldstone, Higgs, Varma, Altland}. The spectrum of the Mott phase 
shows two gapped modes, whereas the superfluid phase has a gapless 
Goldstone mode and a gapped mode, which in condensed matter physics 
is called Higgs mode \cite{Varma}. These features have been recently 
observed experimentally \cite{Endres}. However, an experimental 
investigation of the equation of state around the superfluid-Mott 
phase quantum transition is still missing. 

Near the superfluid-Mott transition the Bose-Hubbard 
model can be mapped into an low-energy and low-momenta effective 
action \cite{India,Stoof,Dupuis,kennett}. 
This action is a generalization of the 
familiar high-temperature Ginzburg-Landau functional \cite{Ginzburg} 
and it contains also time derivatives of the order parameter. These additional 
terms, which make the action formally relativistic, are indeed crucial 
at low temperature. { This kind of effective action has been 
recently used to study the Higgs mode in the BCS-BEC crossover with 
s-wave fermion superfluids \cite{BLiu}}. 

In this work, we adopt this effective relativistic Ginzburg-Landau 
action and, within a functional interation formalism \cite{Altland}, 
we compute the elementary excitations which are in good agreement 
with experimental results  \cite{Endres} and 
share formal analogies with the ones of both the 
non-relativistic and relativistic weakly interacting gases. 
Furthermore, in order to compute the equation 
of state, we regularize divergent integrals by using dimensional 
regularization \cite{t'Hooft,Capper,Salasnich-Toigo}. 
In this way we find that, { within our beyond-mean-field 
Gaussian scheme}, the superfluid-Mott 
phase transition changes from second order to first order 
due to quantum fluctuations. Only at the critical points 
the quantum phase transition remains of the second order. 

Historically, the phenomenon of a first-order transition induced 
by quantum fluctuations was suggested 
by Coleman and Weinberg \cite{Coleman} studying a massless 
charged meson coupled to the electrodynamic field, 
and by Halperin, Lubensky, and Ma \cite{Halperin} 
investigating the fluctuations of the 
electromagnetic field in the superconductor to normal metal transition. 
More recently, this phenomenon has been theoretically predicted 
also for other other phase transitions. For instance, 
the ferromagnet-helix transition in an isotropic quantum Heisenberg 
ferromagnet \cite{Rastelli}, the ferromagnetic-paramagnetic transition 
in local Fermi liquid \cite{Bedell}, and the superfluid-Mott transition 
at the critical points for a two-species bosonic system 
in a three-dimensional optical lattice \cite{Liu}. 
In this paper we are proposing a much more sophisticated effect: 
the { Gaussian} quantum fluctuations of a single $U(1)$ order parameter 
can trigger a quantum phase transition from second- 
to first-order, without the coupling to other dynamical 
fields. Quite remarkably, our theoretical results for the zero-temperature 
equation of state can be directly tested with the 
experimental setups \cite{Greiner,Endres} of 
ultracold alkali-metal atoms loaded into three-dimensional or 
quasi two-dimensional optical lattices. 

\section{Mean-field phase diagram and effective action}

The Bose-Hubbard Hamiltonian is given by
\beq
\label{BH_Hamil}
\hat{H} = - J\sum_{\langle ij \rangle}\hat{a}_{i}^{\dagger}\hat{a}_{j} +
{U\over 2}
\sum_{i}\hat{n}_{i}(\hat{n}_{i}-1) - (\mu -\epsilon)\sum_{i}\hat{n}_{i} \; ,
\eeq
where $\hat{a}_{i}$ is the bosonic annihilation
operator at the site $i$, $\hat{n}_{i}={\hat a}^+_{i} {\hat a}_i$
is the corresponding bosonic number operator, $\epsilon$ is the on-site
energy of bosons, $\mu$ is the chemical potential, $J$ is the hopping term,
which describes the tunneling energy of particles, and $U$ is the on-site
interaction strength of bosons.

\begin{figure}[b]
\centering
\includegraphics[scale=0.48]{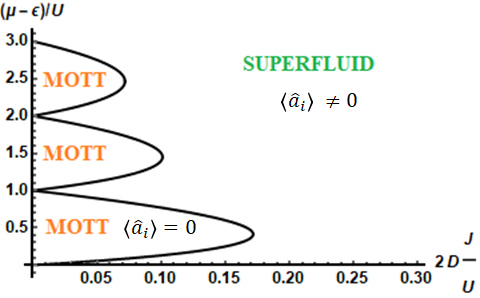}
\caption{Mean-field phase diagram of the 
Bose-Hubbard model at zero temperature in $D$ spatial dimensions. 
$J$ is the hopping energy, $\epsilon$ is the on-site energy, 
$U$ is the on-site interaction energy, $\mu$ is the chemical potential. 
The tips of the Mott lobes are critical points.} 
\label{phase_diagram}
\end{figure}

Within the mean-field decopling approximation \cite{Sachdev} 
the boundary between the superfluid phase and the Mott phase in the 
Bose-Hubbard model are obtained from the equation 
\beqa 
(\mu - \epsilon)^2 - (\mu - \epsilon) \left( U (2 n -1) - 2D J \right) 
\nonumber 
\\
+ 2 D J U + U^2 n (n -1) = 0 \; . 
\label{bordo}
\eeqa
In Fig. \ref{phase_diagram} we plot the 
superfluid-Mott phase diagram of the Bose-Hubbard model at zero temperature, 
obtained from Eq. (\ref{bordo}). Inside the lobes there is the Mott phase, 
characterized by an integer filling number $n$ and where the expectation value 
of the annihilation operator for each site is equal to zero. 
Outside the Mott lobes there is the superfluid phase, 
characterized by a non-vanishing expectation value 
of the annihilation operator for each site. 
The tips of the lobes are critical points. 
The corresponding critical transitions, which occur for $n=1$ at 
$[2DJ/U]_c=0.172$ and $[(\mu-\epsilon)/U]_c = 0.414$, can be controlled 
by varying $U$ at fixed $J$ or varying $J$ at fixed $U$. 
Non-critical transitions, which are the ones not occurring at the tips, 
can be instead obtained, at fixed $J$ and $U$, by changing the 
effective chemical potential $\mu -\epsilon$. 

In the vicinity of the transition line, the Bose-Hubbard model 
can be mapped into the following effective Ginzburg-Landau 
action in Euclidean space \cite{India,Stoof,Dupuis} 
\beqa
\label{eff_act}
S &=& \beta\mathcal{E}_{0} + \int_{0}^{\beta} d\tau \, \int_{V}d^{D}\vec{r}\, 
\Big\{
K_{1}\psi^{*}\frac{\partial}{\partial \tau}\psi + K_{2}\left\vert 
\frac{\partial}{\partial \tau}\psi\right\vert^{2} 
\nonumber 
\\
&+& 
K_{3}\vert\vec{\nabla}\psi\vert^{2} + 
c_{2}\vert \psi \vert^{2} + c_{4}\vert \psi \vert^{4} \Big\} \; , 
\eeqa
where $\psi({\vec x},\tau) = \langle {\hat a}_i(\tau) \rangle$ 
is the space-time dependent order parameter, 
which corresponds to the expectation value of the annihilation operator 
${\hat a}_i(\tau)$ at the imaginary time $\tau$ and at the site $i$ 
associated to spatial position ${\vec x}$. 
Here $\beta=1/T$ with $T$ the absolute temperature, 
$V$ is the volume, and 
\beq 
\mathcal{E}_{0} = V \left( - (\mu - \epsilon) n + {U\over 2} n (n - 1) \right) 
\eeq
is the mean-field energy of the bosonic system in the Mott phase, 
with $n$ the integer filling number of the Mott lobe. 
The parameters $K_1$, $K_2$, $K_3$, $c_2$, and $c_4$ 
of the effective action (\ref{eff_act}) can be expressed 
in terms of the Bose-Hubbard parameters $J$, $U$, and $\mu-\epsilon$, 
and the condition $c_2 = 0$ is equivalent to Eq. (\ref{bordo}). 
We stress that $c_{4}$, $K_{2}$ and $K_{3}$ are 
always positive, while $K_{1}$ vanishes for transitions at the critical 
points (i.e. tips of the lobes). See Appendix 1 for details. 

The effective action (\ref{eff_act}) is a 
generalization of the Ginzburg-Landau functional \cite{Ginzburg} 
and the term which contains $K_2$ makes the action formally relativistic. 
In general, the time-dependent terms are very important for 
an accurate description of the bosonic system at low temperature. 
We shall show that the quantum fluctuations extracted 
from the effective action (\ref{eff_act}) crucially depend 
on $K_1$ and $K_2$ and strongly affect the zero-temperature 
equation of state. In the spirit of the Ginzburg-Landau approach, 
$K_1$, $K_2$, $K_3$ and $c_4$ are calculated at the chosen transition 
point, while $c_2$ remains the only quantity that is tuned by the 
selected control parameter across the transition point. 

\section{Partition function and elementary excitations}

By using the path integral formalism \cite{Altland}, the partition function 
of the bosonic system near the superfluid-Mott transition is then given by 
\beq
Z = \int\,  D[\psi,\psi^{*}] \ \exp\lbrace -S[\psi,\psi^{*}] \rbrace \; . 
\eeq 
From the partition function we can compute the grand-canonical potential as
\beq
\Omega = -\frac{1}{\beta}\ln(Z) \; . 
\label{ovvio}
\eeq
Since our system is homogeneous, the pressure is simply given by 
$P = -\Omega/V$. We write the order parameter as 
\beq
\psi({\vec x},\tau) = \psi_0 + \eta({\vec x},\tau) \; , 
\label{mio}
\eeq
where $\psi_0$ is the uniform and constant order parameter 
and $\eta({\vec x},\tau)$ takes into account space-time fluctuations 
around $\psi_0$. In this way, the grand-canonical potential can be 
written as 
\beq
\Omega = \Omega^{(MF)} + \Omega^{(G)} \; , 
\eeq
where $\Omega^{(MF)}$ is the mean-field grand potential associated $\psi_0$ 
while $\Omega^{(G)}$ is associated to the fluctuating field 
$\eta({\vec x},\tau)$ at the Gaussian level. 

From Eqs. (\ref{eff_act}), (\ref{ovvio}) and (\ref{mio}), 
the mean-field contribution reads 
\beq
\Omega^{(MF)} = {\cal E}_0 + V \ \left( c_2 
\ |\psi_0|^2 + c_4 \ |\psi_0|^4 
\right) \; . 
\eeq
By minimizing this grand potential we obtain, assuming $\psi_0$ real, 
\beq
\psi_0 = \left \{
  \begin{array}{ll}
  0 & \mbox{ if } c_2 > 0 \\
  \sqrt{|c_2| \over 2 c_4}  
    & \mbox{ if } c_2 < 0 
  \end{array}\right.
\eeq
Thus, the mean-field grand-canonical potential becomes 
\beq
\Omega^{(MF)} = \mathcal{E}_{0} - V\frac{c_2^2}{c_{4}} \Theta(-c_2) \; , 
\eeq
where $\Theta(x)$ is the Heaviside step function. We remind 
that the phase diagram shown in Fig. \ref{phase_diagram} is obtained within 
this mean-field picture. 

The zero-temperature Gaussian grand potential is instead given by 
the zero-point energy \cite{Altland,Salasnich-Toigo} 
\beq
\Omega^{(G)} = {1\over 2} \sum_{\vec{q}}\sum_{j=1,2} E_{\vec{q},j} \; , 
\eeq
where $E_{\vec{q},j}$ are the elementary excitations characterized 
by two branches ($j=1,2$). Here $E_{\vec{q},j}=
\omega_j(\vec{q})$, where the frequencies $\omega_j(\vec{q})$ are 
derived from $\mbox{det}[{\cal M}({\bf q},\omega)]=0$. The $4\times 4$ 
matrix ${\cal M}(\vec{q},i\Omega_m)$ is the inverse propagator of 
Gaussian fluctuations with $\vec{q}$ the $D$-dimensional wavevector,  
$\Omega_m=2\pi m/\beta$ the Matsubara frequencies, and $i=\sqrt{-1}$ the 
imaginary unit. See \cite{Faccioli-Salasnich} for details on the derivation of 
${\cal M}(\vec{q},i\Omega_m)$ in the case of both non-relativistic 
and relativistic bosonic actions. 

Inside the Mott phase ($c_2>0$) the energy spectrum reads
\beq
\label{spec_dis}
E_{\vec{q},j} = \sqrt{\frac{K_{3}}{K_{2}}q^{2} + \left( \frac{K_{1}^{2}}
{4K_{2}^{2}}
+ \frac{c_2}{K_{2}} \right)} + 
(-1)^{j}\frac{\vert K_{1}\vert}{2K_{2}} \; , 
\eeq
whereas for the superfluid phase ($c_2<0$) we have
\beqa
E_{\vec{q},j}^2 &=&  \frac{K_{3}}{K_{2}}q^{2} + \left( 
\frac{K_{1}^{2}}{2K_{2}^{2}}- \frac{c_2}{K_{2}} \right) 
\nonumber 
\\
&+&(-1)^{j}\sqrt{
\frac{K_{1}^{2}K_{3}}{K_{2}^{3}}q^{2} + \left( \frac{K_{1}^{2}}{2K_{2}^{2}}
- \frac{c_2}{K_{2}} \right)^{2}} \; . 
\label{spec_ord}
\eeqa
We observe that both modes in the Mott phase are gapped, whereas in the 
superfluid phase we have a gapped (Higgs) mode and a gapless (Goldstone) one 
as expected by Goldstone theorem. { Clearly, for $K_1=0$ the gapless 
mode is linear, namely 
\beq 
E_{\vec{q},1} = \sqrt{K_3\over K_2} \ q \; . 
\eeq 
For $K_1 \neq 0$ 
the small-momentum expansion is a bit more involved but one finds that 
the gapless mode is still linear at small momentum, i.e. 
\beq 
E_{\vec{q},1} \simeq \sqrt{{K_3\over K_2}-{K_1^2 K_3 \over 
2K_2^3 (\frac{K_{1}^{2}}{2K_{2}^{2}} - 
\frac{c_2}{K_{2}})}} \ q \; . 
\eeq
}

\begin{figure}[t]
\centering
\includegraphics[scale=0.48]{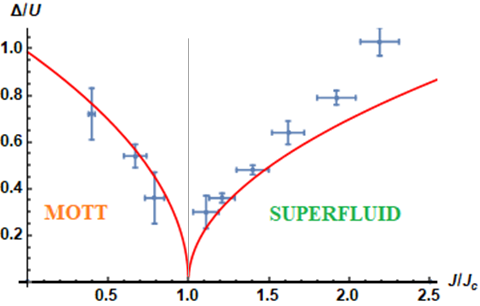}
\caption{We compare the predictions of our theory (solid line) 
for the sum $\Delta$ of the two gaps with experimental data 
(squares with error bars) of Ref. \cite{Endres}. In the horizontal axis $J_c$ 
is the critical value of $J$ at which the critical transition 
of the $n=1$ Mott lobe occurs. 
Within our theoretical scheme $J_c$ is given by $J_c=0.043\ U$.}
\label{data}
\end{figure}

We show in Fig. \ref{data} a comparison between the predictions of our theory
for the sum of the two gaps 
\beq
\Delta=  E_{\vec{q}=\vec{0},1} + E_{\vec{q}=\vec{0},2} \; , 
\eeq
and the available experimental data \cite{Endres}. These data are 
obtained with ultracold and dilute bosonic atoms 
loaded into a quasi two-dimensional optical lattice, 
studying the superfluid-Mott transition at 
the critical point ($K_{1}=0$) of the $n=1$ lobe. 
In the experiment the on-site interaction strength $U$ 
was fixed and the hopping energy $J$ was changed. The figure shows that 
the gaps of our elementary excitations above the uniform 
and constant order parameter are in quite good agreement 
with experimental data for the Mott phase, 
and also with the results in the superfluid phase close to 
the transition line. { Notice that, in the superfluid phase 
within our Gaussian approach we get 
\beq 
\Delta = E_{\vec{q}=\vec{0},2} = 
\sqrt{2} 
\sqrt{\frac{K_{1}^{2}}{2K_{2}^{2}} - \frac{c_{2}}{K_{2}}} \; , 
\eeq
while in the Mott phase we obtain 
\beq 
E_{\vec{q}=\vec{0},j} = \sqrt{\frac{K_1^2}{4K_2^2} 
+ \frac{c_2}{K_2}} + (-1)^j \frac{\vert K_1 \vert}{2K_2} \; . 
\eeq
}
{ When the linear time derivative is 
absent ($K_{1}=0$), at the critical transition ($c_2=0$) all 
the gaps vanish.} We stress that our results are also consistent 
with other theoretical approaches to the elementary excitations 
of the two-dimensional Bose-Hubbard model \cite{Altman,Huber}. 

\begin{figure}[t]
\centering
\includegraphics[scale=0.55]{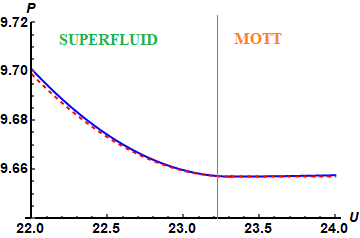} \\
\vskip 0.3cm
\includegraphics[scale=0.55]{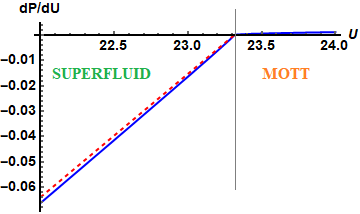} \\
\vskip 0.3cm
\includegraphics[scale=0.55]{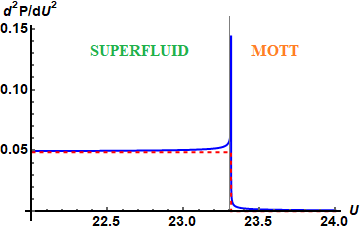}
\caption{Pressure $P=-\Omega/V$ and its derivatives with respect 
to the control parameter $U$, across the 
critical point (tip) of the $n=1$ Mott lobe, as a function of $U$. 
Red dashed line: mean-field theory. Blue solid line: 
beyond-mean-field theory. 
Results obtained for the two-dimensional bosonic system, with $J=1$ and 
$[(\mu -\epsilon)/U]_c=0.414$.}
\label{press_crit}
\end{figure}

\section{Beyond mean-field equation of state} 

We now compute the beyond mean-field equation of state. First of all, 
we consider the case with a vanishing linear time derivative 
of the order parameter ($K_{1}=0$). We have seen that this 
very special case corresponds to a transition across the critical points 
(tips of the Mott lobes). In this case, the Gaussian correction 
to the mean-field grand-canonical potential reads 
\beq
\Omega^{(G)} = \frac{1}{1+\Theta(-c_2)} 
\sum_{j=1,2} \sum_{\vec{q}} 
\sqrt{K_3 q^2 + |c_2| 
+ (-1)^j |c_2|\Theta(-c_2) } \; . 
\eeq
After performing the continuum limit in the sum over momenta 
and the dimensional regularization of the divergent integrals 
(see Appendix 2 for details), the beyond-mean-field 
grand-canonical potential, that is the zero-temperature 
equation of state, for $D=2$ reads 
\beq
\label{crit_d2}
\Omega = \mathcal{E}_0 - V\frac{c_2^2}{c_4}\Theta(-c_2) 
- V\frac{|c_2|^{3/2}}{3\pi K_3} 
\frac{1}{\sqrt{1+\Theta(-c_2)}} \; , 
\eeq
while for $D=3$ it is given by 
\beqa
\Omega &=& \mathcal{E}_0 - V \frac{c_2^2}{c_4} \Theta(-c_2) 
- V\frac{c_2^2}{16\pi^2  K_3^{3/2}} 
\nonumber 
\\
&\cdot& \Big[ 
\ln\left( \frac{q_0K_3^{1/2}}{|c_2|^{1/2}} \right) 
+ \frac{3}{4} - \frac{\gamma}{2} \Big] [1+\Theta(-c_2)] \; , 
\label{crit_c3}
\eeqa
where $\gamma=0.577$ is the Eulero-Mascheroni constant and $q_{0}$ is an 
ultraviolet cut-off in the momenta, related to the maximal length scale 
of the system.  

In Fig. \ref{press_crit} we report our predictions 
for the pressure $P=-\Omega/V$ in the case of a critical 
transition ($K_1=0$). In the figure we consider a two-dimensional 
system ($D=2$) and the superfluid-Mott transition at the tip 
of the $n=1$ Mott lobe. We plot the pressure $P$ 
as a function of the control parameter $U$. 
Fig. \ref{press_crit} shows that the first-order derivative of the 
pressure with the respect to $U$ is 
continuous. Instead, the second-order derivative 
of the pressure with the respect to $U$ has a divergence 
at the transition. Remarkably, this case with $D=2$ is analogous to what 
is found for the specific heat of a $D=3$ system 
in the classical Ginzburg-Landau theory \cite{Kardar}. 
In the case $D=3$ we find similar results. Thus, at the critical 
points, the superfluid-Mott phase transition is of 
the second order both in $D=2$ and $D=3$, and the 
Gaussian quantum fluctuations do not change the order of the transition. 

\begin{figure}[t]
\centering
\includegraphics[scale=0.55]{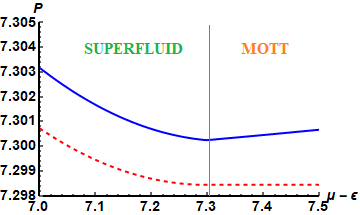} \\
\vskip 0.3cm
\includegraphics[scale=0.55]{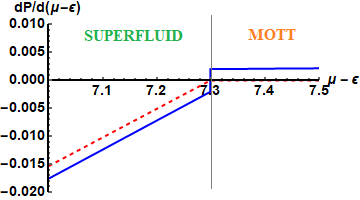} \\
\vskip 0.3cm
\includegraphics[scale=0.55]{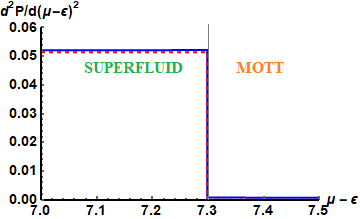}
\caption{Pressure $P=-\Omega/V$ and its derivatives with respect 
to the control parameter $\mu-\epsilon$, 
across a non-critical transition point of the $n=1$ Mott lobe, 
as a function of $\mu-\epsilon$. 
Red dashed line: mean-field theory. Blue solid line: beyond-mean-field theory. 
Results obtained for the two-dimensional bosonic system, 
with $J=1$ and $U=25$.}
\label{press_noncrit}
\end{figure} 

Including the linear time derivative of the order parameter 
($K_1 \neq 0$) the situation changes dramatically. 
In particular we find that for $D=2$ the equation of state reads 
\beq
\label{notcrit_d2}
\Omega = \mathcal{E}_{0} - V\frac{c_2^{2}}
{c_{4}}\Theta(-c_2) - V\frac{\left( \frac{K_{1}^{2}}{4K_{2}} + 
\vert c_2\vert \right)^{\frac{3}{2}}}
{6\pi K_{3}} \; , 
\eeq
while for $D=3$ the grand-canonical potential is given by 
\beqa
\Omega &=& \mathcal{E}_{0} - V\frac{c_2^{2}}{c_{4}} \Theta(-c_2) 
- V\frac{\Big( \frac{K_{1}^{2}}{4K_{2}} 
+ \vert c_2\vert \Big)^{2}} {16\pi^{2} K_{3}^{\frac{3}{2}}}
\nonumber 
\\
&\cdot &\left\{\ln\left[\
\frac{q_{0}K_{3}^{\frac{1}{2}}}{\left( \frac{K_{1}^{2}}{4K_{2}} + 
\vert c_2\vert \right)}\right] + \frac{3}{4} - \frac{\gamma}{2} \right\} \; . 
\label{notcrit_d3}
\eeqa
In both cases, we have that the first-order derivative with the respect
to the control parameter $\mu-\epsilon$ has a jump discontinuity. 
Hence, in contrast to the second-order phase transition predicted 
by the mean-field we have a prediction, 
with the inclusion of the quantum fluctuations, of a first-order phase 
transition. These effects are clearly shown in 
Fig. \ref{press_noncrit}, where we 
plot the behavior of the pressure $P=-\Omega/V$ and its derivatives 
in the case of a non critical transition. 
In the figure we consider again a two-dimensional bosonic 
system ($D=2$) and the superfluid-Mott transition across a non critical 
point of the $n=1$ Mott lobe. Here the chosen 
control parameter is $\mu-\epsilon$ and the derivatives 
of the pressure $P$ are calculated with respect to it, at fixed 
hopping $J$ and on-site interaction strength $U$.

In conclusion, { Gaussian} quantum fluctuations 
strongly modify the properties of the grand-canonical potential 
(or equivalently the pressure) near the superfluid-Mott transition. 
For critical transitions at the tips of the Mott 
lobes, we find a divergent second-order derivative, which is analogous 
to what is found for the specific heat in the classical 
Ginzburg-Landau theory. In all the other non-critical transition points, 
after including the beyond-mean-field Gaussian correction 
a finite discontinuity appears in the first derivative, 
which corresponds to a first-order phase transition. 
Thus, { Gaussian} quantum fluctuations of the order parameter itself 
have a crucial role on the order of the quantum phase 
transition. Our calculations are based on the Bose-Hubbard model but 
they are valid, { at the Gaussian level}, 
for any quantum phase transition described by a Ginzburg-Landau 
action which contains both non-relativistic and 
relativistic time-dependent terms. 

\section*{Acknowledgements}

The authors thank F. Baldovin, M. Baiesi, G. Gradenigo, P.A. Marchetti, 
E. Orlandini, A. Stella, F. Toigo, and A. Trovato for fruitful discussions. 
L.S. acknowledges for partial support the FFABR grant of 
Italian Ministry of Education, University and Research. 

%

%
%
%

\section*{Appendix 1. Parameters of the effective action} 

The parameters of the effective relativistic 
action (\ref{eff_act}) can be written in terms of the Bose-Hubbard 
parameters as \cite{Dupuis}
\beqa 
c_2 &=& \frac{[Un-(\mu-\epsilon)][(\mu-\epsilon)-U(n-1)]}{[(\mu-\epsilon)+U]}
\nonumber 
\\
&-& \frac{2DJ[(\mu+\epsilon)+U]}{[(\mu-\epsilon)+U]} \; , 
\\
K_1 &=& - {\partial c_2\over \partial (\mu -\epsilon)} \; , 
\\
K_2 &=& - {1\over 2} {\partial^2 c_2\over \partial (\mu -\epsilon)^2} \; , 
\\
K_3 &=& J \; , 
\eeqa
where $n$ is the integer filling number, which characterizes the 
lobes of the Mott phase. The quartic coefficient is instead given by 
\beqa 
G^4 c_4 &=& \frac{(n+1)(n+2)}{(2(\mu -\epsilon)
-(2n+1)U)(Un-(\mu -\epsilon))^2} 
\nonumber 
\\
&+& \frac{n(n-1)}{((\mu -\epsilon)-U(n -1))^2(U(2n-3)-2(\mu -\epsilon))} 
\nonumber 
\\ 
&-&\frac{n(n+1)}{((\mu -\epsilon)-Un)(-(\mu -\epsilon)+U(n-1))^2} 
\nonumber 
\\
&-& \frac{n(n+1)}{((\mu -\epsilon)-Un)^2(-(\mu -\epsilon)+U(n-1))} 
\nonumber 
\\
&-& \frac{n^2_0}{(-(\mu -\epsilon)+U(n-1))^3} 
- \frac{(n+1)^2}{((\mu -\epsilon)-Un)^3} \; , 
\eeqa
where 
\beq
G = \frac{n+1}{[(\mu-\epsilon)-Un]} - \frac{n}{[(\mu-\epsilon) - U(n-1)]} \; .
\eeq

\section*{Appendix 2. Dimensional regularization} 

In order to find a finite 
expression for the beyond-mean-field grand potential we adopt 
the dimensional regularization technique \cite{t'Hooft,Capper,Salasnich-Toigo}. 
Here we perform the dimensional 
regularization for $D=2$ and $D=3$, when both the linear and quadratic 
time derivatives are present, i.e. the case $K_{1}\neq 0$ and $K_2\neq 0$. 
The other cases are very similar. First of all, we make the continuum limit 
in the sum over momenta
\beqa
\Omega^{(G)}&=&\sum_{\vec{q}}\sqrt{K_{3}q^{2}+\left(\frac{K_{1}^{2}}
{4K_{2}}+|c_2| \right)} 
\nonumber 
\\
&=& \frac{V}{(2\pi)^{D}}\int\,d^{D}\vec{q}\,\sqrt{K_{3}q^{2}+
\left(\frac{K_{1}^{2}}{4K_{2}}+|c_2| \right)} \; . 
\eeqa 
We now write the integral in the polar coordinates
\beq
\Omega^{(G)}=\frac{2V}{(4\pi)^{\frac{D}{2}}\Gamma(\frac{D}{2})}
\int_{0}^{\infty}\,dq\,\sqrt{K_{3}q^{2}+\left(\frac{K_{1}^{2}}
{4K_{2}}+|c_2| \right)} \; , 
\eeq
where $\Gamma(x)$ is the Euler gamma function. After the transformation 
\beq
Q= \frac{K_{3}}{\left(\frac{K_{1}^{2}}{4K_{2}}+
|c_2|\right)}q^{2}
\eeq
we get
\beq
\Omega^{(G)}=V\frac{\left(\frac{K_{1}^{2}}{4K_{2}} + |c_2|
\right)^{\frac{D+1}{2}}}{(4\pi)^{\frac{D}{2}}
K_{3}^{\frac{D}{2}}
\Gamma(\frac{D}{2})} \int_{0}^{\infty}\,dQ\,Q^{\frac{D}{2}-1}\sqrt{1+Q} \; . 
\eeq
In order to compute this divergent integral, we shift the 
dimensionality as follows 
\beq
D \quad \rightarrow \quad D - \varepsilon \; , 
\eeq
where $\varepsilon$ is a small, complex parameter. We obtain:
\beq
\Omega^{(G)}=V\frac{\left(\frac{K_{1}^{2}}{4K_{2}} + |c_2|
\right)^{\frac{D-\varepsilon+1}{2}}\kappa^{\varepsilon}}
{(4\pi)^{\frac{D-\varepsilon}{2}}
K_{3}^{\frac{D-\varepsilon}{2}}
\Gamma(\frac{D-\varepsilon}{2})}
\int_{0}^{\infty}\,dQ\,Q^{\frac{D-\varepsilon}{2}-1}\sqrt{1+Q} \; , 
\eeq
where we have introduced, for dimensional reason, a momentum scale $\kappa$.
Now, the integral can be written as 
\beq
\int_{0}^{\infty} Q^{\frac{D-\varepsilon}{2}-1}\sqrt{1+Q} =
\frac{\Gamma(\frac{D-\varepsilon}{2})\Gamma(-\frac{(D+1-\varepsilon)}{2})}
{\Gamma(-\frac{1}{2})} \; . 
\eeq
Hence, we obtain for the Gaussian term  
\beq
\Omega^{(G)} = V\frac{\left(\frac{K_{1}^{2}}{4K_{2}} + |c_2| 
\right)^{\frac{D-\varepsilon+1}{2}}\kappa^{\varepsilon}}
{(4\pi)^{\frac{D-\varepsilon}{2}}
K_{3}^{\frac{D-\varepsilon}{2}}}
\frac{\Gamma(-\frac{D}{2}-\frac{1}{2}+\frac{\varepsilon}{2})}
{\Gamma(-\frac{1}{2})} \; . 
\eeq
For $D=2$, we get for $\varepsilon\rightarrow 0$ 
\beq
\Omega^{(G)} = V\frac{\left(\frac{K_{1}^{2}}{4K_{2}} + |c_2| 
\right)^{\frac{3}{2}}} {6\pi K_{3}} \; . 
\eeq
For $D=3$, after regularization we still find a divergent result, because
the Gamma function has poles for negative integers. To isolate 
the divergence, we expand the Gamma function around 
$\varepsilon =0$ 
\beq
\Gamma\left( -2 + \frac{\varepsilon}{2} \right) = \frac{1}{\epsilon} - 
\frac{\gamma}{2} + \frac{3}{4} + O(\varepsilon) \; , 
\eeq
and we get 
\beqa
\frac{\left(\frac{K_{1}^{2}}{4K_{2}} + |c_2|
\right)^{\frac{4-\varepsilon}{2}}\kappa^{\varepsilon}}
{(4\pi)^{\frac{1-\varepsilon}{2}}
K_{3}^{\frac{1-\varepsilon}{2}}}
\frac{\Gamma(-1+\frac{\varepsilon}{2})}{\Gamma(-\frac{1}{2})} 
=\frac{\left(\frac{K_{1}^{2}}{4K_{2}} + |c_2| \right)^{2}}
{(4\pi)^{\frac{3}{2}}K_{3}^{\frac{3}{2}}
\Gamma(-\frac{1}{2})}
\nonumber
\\
\cdot 
\left\{
\frac{3}{4} - \frac{\gamma}{2}- \ln\left[\frac{K_{3}^{\frac{1}{2}}q_{0}}
{\left(\frac{K_{1}^{2}}{4K_{2}} + |c_2|\right)}\right]
\right\} + O(\epsilon) \; , 
\eeqa
where $q_{0}$ is given by
\beq
q_{0} = 2\pi^{\frac{1}{2}}\kappa \; . 
\eeq
Using the above results, we finally obtain the regularized 
Gaussian correction:
\beq
\Omega^{(G)} = - V\frac{\left( \frac{K_{1}^{2}}{4K_{2}} + |c_2|\right)^{2}}
{32\pi^{2} K_{3}^{\frac{3}{2}}}\left\{\ln\left[\
\frac{q_{0}K_{3}^{\frac{1}{2}}}{\left( \frac{K_{1}^{2}}{4K_{2}} + 
|c_2| \right)}\right] - \frac{\gamma}{2} + \frac{3}{4} \right\} \; , 
\eeq
where we have used $\Gamma\left(-\frac{1}{2}\right) = -2\sqrt{\pi}$.

\end{document}